\PassOptionsToPackage{sort&compress}{natbib}
\documentclass[final,3p,times]{elsarticle}
\usepackage[T1]{fontenc}
\usepackage{amsmath,amssymb}
\usepackage{xcolor,etoolbox}
\usepackage{tikz}
\usepackage{placeins}
\usetikzlibrary{arrows.meta}
\usepackage[hidelinks]{hyperref}
\journal{Journal of Subatomic Particles and Cosmology}

\makeatother
\definecolor{chiralred}{RGB}{180,45,45}
\definecolor{axialgreen}{RGB}{30,120,65}
\hypersetup{pdfauthor={Dan Zhang, Heng-Tong Ding, Jose Javier Hernandez Hernandez},pdftitle={Fate of chiral and $U(1)_A$ symmetries in hot QCD matter under background magnetic fields}}

\begin{document}
\begin{frontmatter}
\title{Fate of chiral and $U(1)_A$ symmetries in hot QCD matter under background magnetic fields}
\author[ccnu]{Dan Zhang\fnref{speaker}}
\author[ccnu]{Heng-Tong Ding}
\author[ccnu]{Jos\'e Javier Hern\'andez Hern\'andez}
\fntext[speaker]{Speaker}
\affiliation[ccnu]{organization={Key Laboratory of Quark and Lepton Physics (MOE) and Institute of Particle Physics, Central China Normal University},city={Wuhan},postcode={430079},country={China}}
\begin{abstract}
     We summarize our recent lattice QCD study of chiral and singlet $U(1)_A$ symmetry breaking in background magnetic fields. The magnetic field explicitly reduces the non-singlet flavor symmetry through the unequal light-quark electric charges. We therefore study two neutral-channel susceptibility differences: $\chi_{\pi^0}-\chi_\sigma$ for chiral symmetry breaking and $\chi_{\pi^0}-\chi_{\delta^0}$ for singlet $U(1)_A$ symmetry breaking. The calculation uses fixed-scale $(2+1)$-flavor HISQ ensembles with $m_l=m_s^{\mathrm{phys}}/10$, corresponding to a zero-field pion mass of $M_\pi\simeq220\,\mathrm{MeV}$. Magnetic fields enhance both susceptibility differences at low temperatures. Near the crossover, sufficiently strong fields instead suppress both differences. The suppression sets in at larger magnetic fields and is milder for the singlet-axial susceptibility difference than for the neutral chiral susceptibility difference.
\end{abstract}
\end{frontmatter}

\section{Introduction}
Spontaneous chiral symmetry breaking and the axial anomaly are two of the
central features of nonperturbative QCD~\cite{Adler:1969gk,Bell:1969ts,Fujikawa:1979ay}.
Lattice studies have found that $U(1)_A$-breaking effects can persist
above the chiral crossover~\cite{Ding:2026blh}.
For two massless light quarks at zero magnetic field, the strength
of $U(1)_A$ breaking near the chiral transition can affect its order
and universality class~\cite{Pisarski:1983ms,Pelissetto:2013hqa}.

In an external magnetic field, the unequal electric charges of
the up and down quarks explicitly reduce the non-singlet flavor
symmetry of the light-quark sector.
The field also modifies the dynamics of chiral symmetry breaking.
Lattice QCD studies have found that magnetic fields enhance the
light-quark condensate at low temperature but can suppress it
near the crossover, corresponding to magnetic catalysis and
inverse magnetic catalysis, respectively~\cite{Bali:2011qj,Bali:2012zg,DElia:2018xwo,Ding:2022tqn}.

While the magnetic response of the chiral sector has been
extensively studied on the lattice, that of singlet
$U(1)_A$ breaking and its effective restoration remains
less explored. A pure magnetic background does not generate 
an electromagnetic contribution to the axial anomaly, since
$F_{\mu\nu}^{\rm em}\widetilde F_{\mu\nu}^{\rm em}
\propto \boldsymbol E\cdot\boldsymbol B=0$. 
Nevertheless, magnetic fields can modify observables related
to $U(1)_A$ breaking, including topological fluctuations
and meson-partner correlations.
The magnetic dependence of the topological susceptibility
has been investigated in chiral perturbation
theory~\cite{Adhikari:2021lbl} and lattice
QCD~\cite{Brandt:2024gso}.
The lattice results show enhancement at low temperature and suppression around the crossover relative to the zero-field susceptibility at the same temperature.
In the meson sector, an NJL model study with quark anomalous
magnetic moments found suppression of a $U(1)_A$-sensitive
susceptibility difference near the crossover, a behavior
termed axial inverse magnetic catalysis~\cite{Wang:2021dcy}.

In these proceedings, we present results from our recent work~\cite{Ding:axial2026} on the magnetic responses of chiral and singlet-axial susceptibility differences in the neutral meson sector.
We first identify the neutral partner relations associated
with the reduced flavor symmetry in a magnetic background.
These relations motivate the susceptibility differences
$\chi_{\pi^0}-\chi_\sigma$ and
$\chi_{\pi^0}-\chi_{\delta^0}$ as probes of neutral chiral
and singlet-axial symmetry breaking, respectively.
Using fixed-scale $(2+1)$-flavor HISQ ensembles, we determine their temperature and magnetic-field dependences.

\section{Symmetry partners and observables}
\label{sec:partners}
For the light-quark doublet $q=(u,d)^T$, we identify the symmetry transformations in the two-flavor chiral limit, $m_u=m_d=0$. At $B=0$, the classical flavor symmetry is $SU(2)_L\times SU(2)_R\times U(1)_V\times U(1)_A$. The vacuum spontaneously breaks $SU(2)_L\times SU(2)_R$ to vector isospin $SU(2)_V$, while the quantum anomaly breaks the singlet $U(1)_A$. The vector $U(1)_V$, associated with conservation of net light-quark number, remains unbroken.

\begin{figure}[!htbp]
     \centering
     \begin{tikzpicture}[
         x=1cm,y=1cm,>=Stealth,
         meson/.style={
             font=\large,
             inner sep=4pt
         },
         field/.style={
             font=\normalsize,
             text=black!65
         },
         chiral/.style={
             <->,
             line width=0.9pt,
             chiralred
         },
         axial/.style={
             <->,
             line width=0.9pt,
             axialgreen
         },
         every node/.style={font=\small}
     ]
     \foreach \shift/\field/\pion/\delt/\non in {
         0/{B=0}/{\pi}/{\delta}/{SU(2)_A},
         7.1/{B\ne0}/{\pi^0}/{\delta^0}/{U(1)_A^{(3)}}
     }{
         \begin{scope}[xshift=\shift cm]
             \node[meson] (p) at (0.8,2.0) {$\pion$};
             \node[meson] (s) at (4.0,2.0) {$\sigma$};
             \node[meson] (d) at (0.8,0.0) {$\delt$};
             \node[meson] (e) at (4.0,0.0) {$\eta$};
     
             \node[field] at (2.4,1.0) {$\field$};
     
             \draw[chiral] (p) --
                 node[above=3pt,text=chiralred] {$\non$} (s);
             \draw[chiral] (d) --
                 node[below=3pt,text=chiralred] {$\non$} (e);
     
             \draw[axial] (p) --
                 node[left=3pt,text=axialgreen] {$U(1)_A$} (d);
             \draw[axial] (s) --
                 node[right=3pt,text=axialgreen] {$U(1)_A$} (e);
         \end{scope}
     }
     \end{tikzpicture}
     \caption{Meson partner relations at $B=0$ (left) and in the
     neutral sector at $B\ne0$ (right).
     Horizontal and vertical arrows denote non-singlet and singlet
     axial transformations, respectively.}
     \label{fig:partners}
     \end{figure}
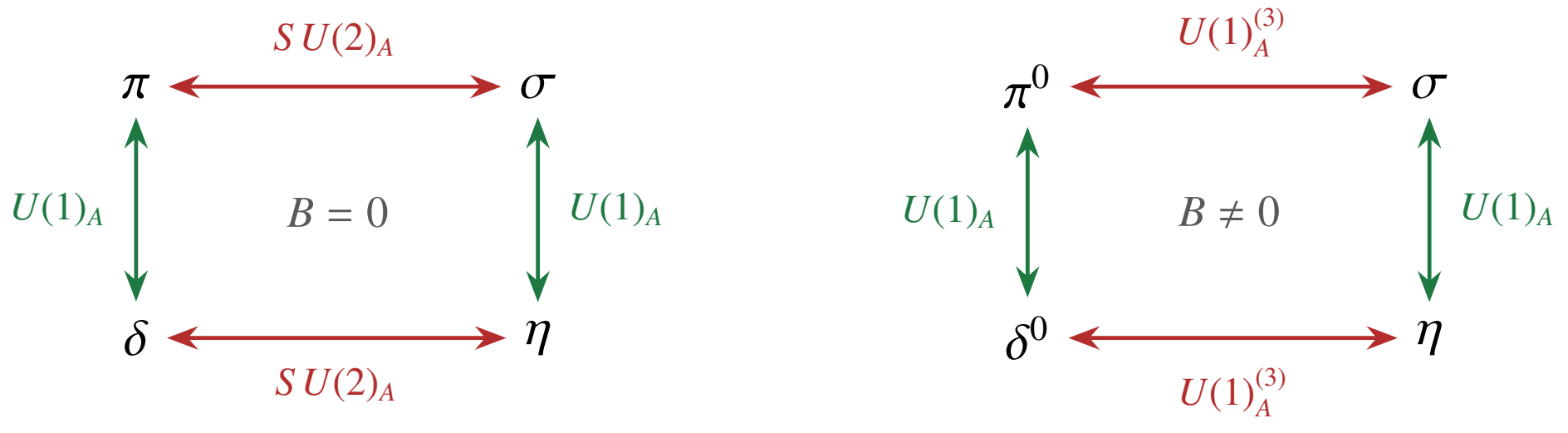

We introduce the scalar and pseudoscalar interpolators
\begin{equation}
\begin{aligned}
\pi^a&=i\bar q\gamma_5\frac{\tau^a}{\sqrt{2}}q,&
\delta^a&=\bar q\frac{\tau^a}{\sqrt{2}}q,\\
\sigma&=\bar q\frac{\mathbf{1}}{\sqrt{2}}q,&
\eta&=i\bar q\gamma_5\frac{\mathbf{1}}{\sqrt{2}}q,
\end{aligned}
\label{eq:operators}
\end{equation}
where $\tau^a$, with $a=1,2,3$, are Pauli matrices in light-flavor space. A non-singlet axial rotation in flavor direction $a$ connects $\pi^a$ with $\sigma$ and $\delta^a$ with $\eta$. Singlet $U(1)_A$ transformations instead connect $\pi^a$ with $\delta^a$ and $\sigma$ with $\eta$. These transformation properties define the axial partners~\cite{HotQCD:2012vvd,Buchoff:2013nra}. At $B=0$, unbroken $SU(2)_V$ makes the correlators identical for the three components of each isovector. We summarize these zero-field partner relations in the left panel of Fig.~\ref{fig:partners}, where $\pi$ and $\delta$ denote representative isovector components, without a sum over $a$.

A magnetic background distinguishes the isovector components through the charge matrix
\begin{equation}
Q=\operatorname{diag}(2/3,-1/3)=\tfrac16\mathbf{1}+\tfrac12\tau^3.
\end{equation}
Since $[Q,\tau^3]=0$ but $[Q,\tau^{1,2}]\ne0$, the magnetic field explicitly reduces the non-singlet vector and axial flavor symmetries to $U(1)_V^{(3)}$ and $U(1)_A^{(3)}$, respectively, in the chiral limit. The superscript $(3)$ labels the flavor generator. For a pure magnetic field the electromagnetic anomaly density vanishes; the singlet gluonic anomaly remains. We summarize the resulting neutral-sector partner relations in the right panel of Fig.~\ref{fig:partners}. The surviving non-singlet axial transformation connects $\pi^0\equiv\pi^3$ with $\sigma$ and $\delta^0\equiv\delta^3$ with $\eta$, while the singlet axial transformation connects $\pi^0$ with $\delta^0$ and $\sigma$ with $\eta$.

Comparing the two panels of Fig.~\ref{fig:partners}, we see that the neutral sector retains the zero-field pattern of partner pairings, while the non-singlet axial transformations are restricted to $U(1)_A^{(3)}$. This contrast motivates our comparison of chiral and singlet $U(1)_A$ symmetry breaking through neutral-channel susceptibility differences. For an interpolator $\mathcal O$, the lattice susceptibility is
\begin{equation}
\chi_{\mathcal O}=\sum_x
\left[\langle\mathcal O(x)\mathcal O(0)\rangle
-\langle\mathcal O\rangle^2\right],
\label{eq:susceptibility}
\end{equation}
where the sum covers all Euclidean lattice sites. Lattice fields and susceptibilities in these expressions are written in lattice units. At $B=0$, the isovector susceptibilities are component-independent and denoted $\chi_\pi$ and $\chi_\delta$, without a sum over components. The corresponding neutral susceptibilities must be distinguished at $B\ne0$. We define the normalized chiral-partner splitting $O_\chi$ and singlet $U(1)_A$ partner splitting $O_A$ as
\begin{align}
O_\chi&=\frac{m_s^2}{f_K^4}(\chi_{\pi^0}-\chi_\sigma),&
O_A&=\frac{m_s^2}{f_K^4}(\chi_{\pi^0}-\chi_{\delta^0}).
\label{eq:splittings}
\end{align}
Here $f_K$ is the zero-field kaon decay constant. The factor $m_s^2/f_K^4$ makes both susceptibility differences dimensionless and cancels the multiplicative renormalization factors associated with the quark bilinears. 
Small values of $O_\chi$ and $O_A$ indicate approximate degeneracy of the respective partner susceptibilities; they do not, by themselves, establish exact symmetry restoration at finite quark mass.

For the numerical evaluation, the scalar susceptibilities separate into quark-line connected and disconnected contractions,
\begin{equation}
\chi_\sigma=\chi_{\rm con}+\chi_{\rm disc}^{(+)},
\qquad
\chi_{\delta^0}=\chi_{\rm con}+\chi_{\rm disc}^{(-)}.
\label{eq:scalar_channels}
\end{equation}
Here $\chi_{\rm con}$ is the flavor average of the connected scalar contractions, evaluated using $\operatorname{Tr}\mathcal M_f^{-2}$. To define the disconnected terms, let $X_f=\operatorname{Tr}\mathcal M_f^{-1}$ be the integrated scalar loop for flavor $f=u,d$, with $\mathcal M_f$ the fermion matrix and the trace over spacetime and internal indices. In the same flavor normalization as Eq.~\eqref{eq:operators},
\begin{equation}
\chi_{\rm disc}^{(\pm)}=\frac{1}{2V_4}
\left[\left\langle(X_u\pm X_d)^2\right\rangle
-\left\langle X_u\pm X_d\right\rangle^2\right],
\label{eq:scalar_disc}
\end{equation}
where $V_4=N_xN_yN_zN_\tau$. Equation~\eqref{eq:scalar_disc} evaluates the disconnected scalar contractions of Eq.~\eqref{eq:susceptibility}. At $B=0$, equal light masses give $X_u=X_d$ on each gauge configuration and hence $\chi_{\rm disc}^{(-)}=0$. At nonzero field this cancellation no longer follows, so the neutral delta susceptibility must include its disconnected term.

The difference between the two partner splittings is
\begin{equation}
O_{\rm disc}\equiv O_A-O_\chi
=\frac{m_s^2}{f_K^4}
\left(\chi_{\rm disc}^{(+)}-\chi_{\rm disc}^{(-)}\right).
\label{eq:Odisc}
\end{equation}

For the neutral-pion channel, we use the integrated non-singlet axial Ward identity~\cite{Ding:2020hxw},
\begin{equation}
2m_l\chi_{\pi^0}=\langle\bar u u\rangle+\langle\bar d d\rangle.
\label{eq:ward}
\end{equation}
The integrated $U(1)_\epsilon$ Ward identity remains exact at finite lattice spacing in the presence of the external $U(1)$ link phases~\cite{Kilcup:1986dg}. It determines the pseudoscalar susceptibility entering both differences directly from the light-quark condensates.

\section{Lattice setup}
\label{sec:lattice_setup}

Our calculations use the gauge ensembles of
Ref.~\cite{Ding:2021cwv}, generated with the HISQ fermion
action~\cite{Follana:2006rc} and a tree-level improved Symanzik gauge action
for $(2+1)$-flavor QCD.
The strange-quark mass is fixed to its physical value,
and the degenerate light-quark masses are set to
$m_u=m_d\equiv m_l=m_s^{\rm phys}/10$.
At zero magnetic field, the pion mass is approximately
$220$ MeV and the kaon decay constant is
$f_K\simeq112.5$ MeV.
The results presented here are obtained on $32^3$ spatial lattices
at a fixed lattice spacing of $a\simeq0.117$ fm.
The temperature is varied by changing $N_\tau$,
with $T=1/(aN_\tau)$, over the range
$T\simeq17$--$281$ MeV.

A uniform magnetic field in the $z$ direction is introduced
through flavor-dependent $U(1)$ phases multiplying the
gauge links in the fermion matrix.
Periodic spatial boundary conditions require
quantized magnetic flux~\cite{DElia:2010abb,Smit:1986fn,Al-Hashimi:2008quu,Bali:2011qj},
\begin{equation}
eB=\frac{6\pi N_b}{N_xN_y}a^{-2},
\qquad N_b\in\mathbb Z.
\end{equation}
We use
$N_b=0,1,2,3,4,6,8,10,12,16,20,24,32,40,48$,
covering magnetic fields up to
$eB\simeq2.51\,\mathrm{GeV}^2$.
Since the lattice spacing and spatial dimensions are
held fixed, a given $N_b$ corresponds to the same
physical magnetic field at all temperatures.

The fermion traces needed for the chiral condensates
and scalar susceptibilities are estimated stochastically
using 51 random vectors per gauge configuration.
The neutral-pion susceptibility is then obtained from
the light-quark condensates using Eq.~\eqref{eq:ward}.

\section{Results and discussion}
\label{sec:results}
\begin{figure}[!htbp]
    \centering
    \includegraphics[width=0.46\linewidth]{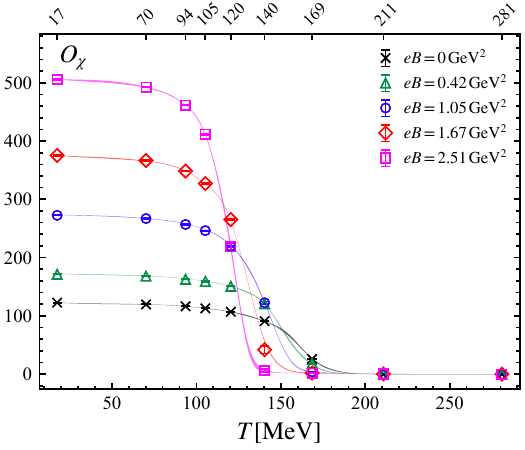}\hfill
    \includegraphics[width=0.46\linewidth]{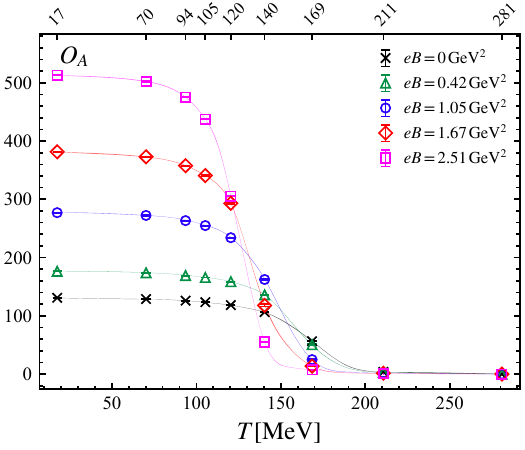}
    \caption{Temperature dependence of the chiral-partner splitting
    $O_\chi$ (left) and the singlet $U(1)_A$ partner splitting $O_A$
    (right) at selected magnetic fields.}
    \label{fig:thermal}
\end{figure}

We first discuss the temperature dependence of the two
partner splittings at fixed magnetic field, shown in
Fig.~\ref{fig:thermal}.
Both splittings decrease with increasing temperature, with a
pronounced reduction between $T\simeq120$ and $169$ MeV at $eB = 0$.
This reduction shifts toward lower temperatures as the
magnetic field increases. At $T\simeq169$ MeV, $O_\chi$ is substantially reduced,
while $O_A$ retains a larger splitting at the same field,
indicating that the neutral chiral partners are closer
to degeneracy than the singlet-axial partners. With further heating to $T\simeq211$ and $281$ MeV, both splittings become small on the scale of
Fig.~\ref{fig:thermal} for all magnetic fields studied. 

\begin{figure}[!htbp]
    \centering
    \includegraphics[width=0.46\textwidth]{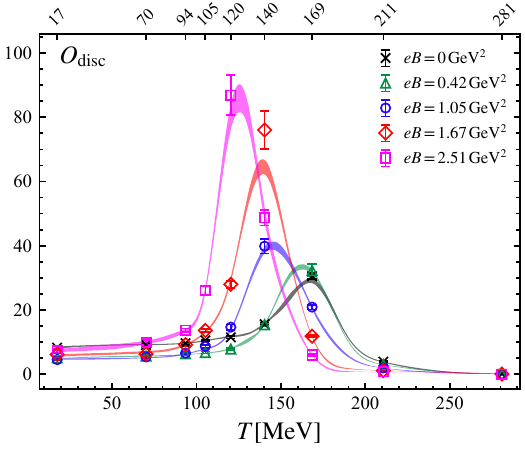}\hfill
    \includegraphics[width=0.46\textwidth]{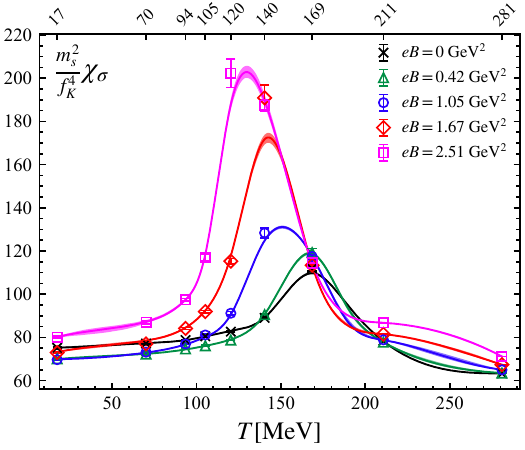}
    \caption{Temperature dependence of the difference between
    the two partner splittings, $O_{\rm disc}=O_A-O_\chi$ (left),
    and the dimensionless full scalar-isoscalar susceptibility
    $m_s^2\chi_\sigma/f_K^4$ (right), at selected magnetic fields.}
    \label{fig:chi_sigma_T}
\end{figure}

The left panel of Fig.~\ref{fig:chi_sigma_T} shows $O_{\rm disc}$, the difference between the two partner
splittings. It is largest at intermediate temperatures,
and becomes small at low and high temperatures.
The temperature region where the two splittings differ most shifts toward lower temperatures as the magnetic field increases.

As seen in the right panel of Fig.~\ref{fig:chi_sigma_T}, our data for $m_s^2\chi_\sigma/f_K^4$ at zero magnetic field suggest $T_{\mathrm{pc}}(0)\simeq170\,\mathrm{MeV}$, consistent with previous estimates~\cite{Ding:2021cwv,Ding:2022tqn}. As the magnetic field increases, the susceptibility enhancement shifts toward lower temperatures, suggesting a decrease in the crossover temperature. The spacing between the simulated temperatures, however, precludes a precise determination of $T_{\mathrm{pc}}(eB)$.

\begin{figure}[!htbp]
\centering
\includegraphics[width=0.46\linewidth]{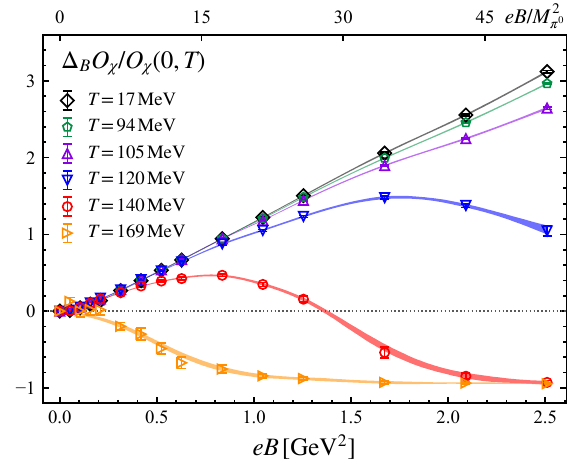}\hfill
\includegraphics[width=0.46\linewidth]{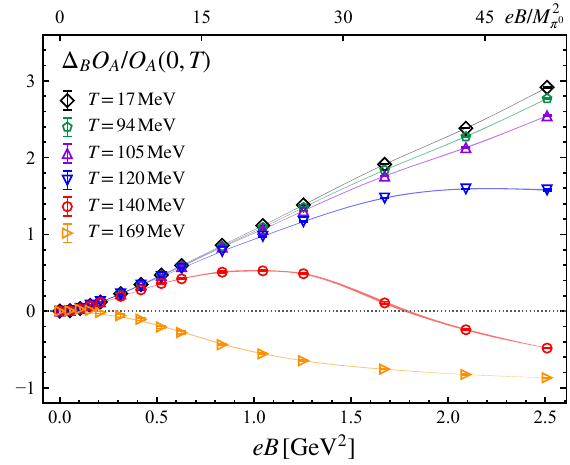}
\caption{Relative magnetic responses $R_{O_\chi}$ (left) and $R_{O_A}$ (right) at selected temperatures.  The dotted line marks $R=0$.}
\label{fig:relative}
\end{figure}

We then compare the magnetic responses below and near the crossover by normalizing the field-induced change in each splitting to its zero-field value at the same temperature,
\begin{equation}
R_{O_i}(B,T)=\frac{O_i(B,T)-O_i(0,T)}{O_i(0,T)},\qquad i=\chi,A.
\label{eq:relative}
\end{equation}
In Fig.~\ref{fig:relative}, positive and negative
$R_{O_i}$ indicate enhancement and suppression relative
to zero field, respectively.
We restrict this comparison to $T\lesssim169$ MeV because
the small zero-field splittings at higher temperatures
make the ratios sensitive to tiny absolute changes.
The shaded bands show the central 68\% intervals from
resampled one-dimensional smoothing fits in $eB$ at fixed
temperature and serve as guides to the eye.

At low temperatures, we find comparable fractional enhancements in the two channels. The difference becomes pronounced at $T\simeq140$ MeV, where the chiral response turns negative at a smaller field than the singlet-axial response. At the strongest field, $R_{O_\chi}$ approaches $-1$, while $R_{O_A}$ is approximately $-0.5$. Thus the chiral-partner splitting is reduced to a small fraction of its zero-field value, whereas the singlet $U(1)_A$ partner splitting retains roughly half of its reference value.

The observable $O_\chi$, which characterizes chiral symmetry breaking in the neutral sector, is enhanced by magnetic fields at low temperatures and suppressed by sufficiently strong fields near the crossover. This pattern is qualitatively similar to the magnetic catalysis and inverse magnetic catalysis observed in the light-quark condensate. For $O_A$, the large-field reduction instead measures a weakening of singlet $U(1)_A$-breaking effects in the neutral partner susceptibilities. We therefore describe it as an inverse-magnetic-catalysis-like response of the singlet $U(1)_A$ partner splitting, qualitatively similar to the axial inverse magnetic catalysis discussed in Ref.~\cite{Wang:2021dcy}.

\section{Conclusions}
We have studied the magnetic response of chiral and singlet $U(1)_A$ symmetry breaking in $(2+1)$-flavor lattice QCD. The reduced non-singlet flavor symmetry in a magnetic field motivates the use of neutral meson partners. Their susceptibility differences, $\chi_{\pi^0}-\chi_\sigma$ and $\chi_{\pi^0}-\chi_{\delta^0}$, probe chiral and singlet $U(1)_A$ symmetry breaking, respectively.

Both partner splittings decrease with increasing temperature, while the singlet $U(1)_A$ partner splitting persists to higher temperatures. At low temperatures, magnetic fields enhance both splittings. Near the crossover, sufficiently strong fields instead suppress them below their zero-field values at the same temperature. The singlet-axial suppression sets in at larger fields and is weaker in fractional terms than the chiral suppression. These results are obtained at a single lattice spacing and a heavier-than-physical pion mass. Calculations at finer lattice spacings and lighter quark masses are needed to quantify discretization and quark-mass effects and determine how these trends evolve toward the continuum and chiral limits.

\section*{Acknowledgments}
We acknowledge support from the National Natural Science Foundation of China (Grants 12293064, 12293060 and 12325508), the National Key Research and Development Program of China (Contract 2022YFA1604900), and the Fundamental Research Funds for the Central Universities, Central China Normal University (Grants 30101250314 and 30106250152). Simulations were performed at the Nuclear Science Computing Center at Central China Normal University ($\mathrm{NSC}^3$) and Wuhan Supercomputing Center.

\bibliographystyle{elsarticle-num}
\bibliography{refs}

\begin{thebibliography}{10}
\expandafter\ifx\csname url\endcsname\relax
  \def\url#1{\texttt{#1}}\fi
\expandafter\ifx\csname urlprefix\endcsname\relax\def\urlprefix{URL }\fi
\expandafter\ifx\csname href\endcsname\relax
  \def\href#1#2{#2} \def\path#1{#1}\fi

\bibitem{Adler:1969gk}
S.~L. Adler, {Axial vector vertex in spinor electrodynamics}, Phys. Rev. 177
  (1969) 2426--2438.
\newblock \href {https://doi.org/10.1103/PhysRev.177.2426}
  {\path{doi:10.1103/PhysRev.177.2426}}.

\bibitem{Bell:1969ts}
J.~S. Bell, R.~Jackiw, {A PCAC puzzle: $\pi^0 \to \gamma \gamma$ in the
  $\sigma$ model}, Nuovo Cim. A 60 (1969) 47--61.
\newblock \href {https://doi.org/10.1007/BF02823296}
  {\path{doi:10.1007/BF02823296}}.

\bibitem{Fujikawa:1979ay}
K.~Fujikawa, {Path Integral Measure for Gauge Invariant Fermion Theories},
  Phys. Rev. Lett. 42 (1979) 1195--1198.
\newblock \href {https://doi.org/10.1103/PhysRevLett.42.1195}
  {\path{doi:10.1103/PhysRevLett.42.1195}}.

\bibitem{Ding:2026blh}
H.-T. Ding, {Axial $U(1)$ Symmetry Breaking in Hot QCD: From Topology to the
  Chiral Phase Transition}, 2026.
\newblock \href {http://arxiv.org/abs/2609.21907} {\path{arXiv:2609.21907}}.

\bibitem{Pisarski:1983ms}
R.~D. Pisarski, F.~Wilczek, {Remarks on the Chiral Phase Transition in
  Chromodynamics}, Phys. Rev. D 29 (1984) 338--341.
\newblock \href {https://doi.org/10.1103/PhysRevD.29.338}
  {\path{doi:10.1103/PhysRevD.29.338}}.

\bibitem{Pelissetto:2013hqa}
A.~Pelissetto, E.~Vicari, {Relevance of the axial anomaly at the
  finite-temperature chiral transition in QCD}, Phys. Rev. D 88~(10) (2013)
  105018.
\newblock \href {http://arxiv.org/abs/1309.5446} {\path{arXiv:1309.5446}},
  \href {https://doi.org/10.1103/PhysRevD.88.105018}
  {\path{doi:10.1103/PhysRevD.88.105018}}.

\bibitem{Bali:2011qj}
G.~S. Bali, F.~Bruckmann, G.~Endrodi, Z.~Fodor, S.~D. Katz, S.~Krieg,
  A.~Schafer, K.~K. Szabo, {The QCD phase diagram for external magnetic
  fields}, JHEP 02 (2012) 044.
\newblock \href {http://arxiv.org/abs/1111.4956} {\path{arXiv:1111.4956}},
  \href {https://doi.org/10.1007/JHEP02(2012)044}
  {\path{doi:10.1007/JHEP02(2012)044}}.

\bibitem{Bali:2012zg}
G.~S. Bali, F.~Bruckmann, G.~Endrodi, Z.~Fodor, S.~D. Katz, A.~Schafer, {QCD
  quark condensate in external magnetic fields}, Phys. Rev. D 86 (2012) 071502.
\newblock \href {http://arxiv.org/abs/1206.4205} {\path{arXiv:1206.4205}},
  \href {https://doi.org/10.1103/PhysRevD.86.071502}
  {\path{doi:10.1103/PhysRevD.86.071502}}.

\bibitem{DElia:2018xwo}
M.~D'Elia, F.~Manigrasso, F.~Negro, F.~Sanfilippo, {QCD phase diagram in a
  magnetic background for different values of the pion mass}, Phys. Rev. D
  98~(5) (2018) 054509.
\newblock \href {http://arxiv.org/abs/1808.07008} {\path{arXiv:1808.07008}},
  \href {https://doi.org/10.1103/PhysRevD.98.054509}
  {\path{doi:10.1103/PhysRevD.98.054509}}.

\bibitem{Ding:2022tqn}
H.~T. Ding, S.~T. Li, J.~H. Liu, X.~D. Wang, {Chiral condensates and screening
  masses of neutral pseudoscalar mesons in thermomagnetic QCD medium}, Phys.
  Rev. D 105~(3) (2022) 034514.
\newblock \href {http://arxiv.org/abs/2201.02349} {\path{arXiv:2201.02349}},
  \href {https://doi.org/10.1103/PhysRevD.105.034514}
  {\path{doi:10.1103/PhysRevD.105.034514}}.

\bibitem{Adhikari:2021lbl}
P.~Adhikari, {Topological susceptibility in a uniform magnetic field}, Phys.
  Lett. B 825 (2022) 136826.
\newblock \href {http://arxiv.org/abs/2103.05048} {\path{arXiv:2103.05048}},
  \href {https://doi.org/10.1016/j.physletb.2021.136826}
  {\path{doi:10.1016/j.physletb.2021.136826}}.

\bibitem{Brandt:2024gso}
B.~B. Brandt, G.~Endr{\H{o}}di, J.~J.~H. Hern{\'a}ndez, G.~Mark{\'o}, {Impact
  of extreme magnetic fields on the QCD topological susceptibility in the
  vicinity of the crossover region}, JHEP 12 (2024) 228, [Erratum: JHEP 03, 034
  (2025)].
\newblock \href {http://arxiv.org/abs/2409.00796} {\path{arXiv:2409.00796}},
  \href {https://doi.org/10.1007/JHEP12(2024)228}
  {\path{doi:10.1007/JHEP12(2024)228}}.

\bibitem{Wang:2021dcy}
Y.~Wang, S.~Matsuzaki, {Axial inverse magnetic catalysis}, Phys. Rev. D 105~(7)
  (2022) 074015.
\newblock \href {http://arxiv.org/abs/2110.10432} {\path{arXiv:2110.10432}},
  \href {https://doi.org/10.1103/PhysRevD.105.074015}
  {\path{doi:10.1103/PhysRevD.105.074015}}.

\bibitem{Ding:axial2026}
H.-T. Ding, J.~J. Hern{\'a}ndez~Hern{\'a}ndez, D.~Zhang, Chiral and {$U(1)_A$}
  symmetries in background magnetic fields from lattice {QCD} (2026).
\newblock \href {http://arxiv.org/abs/2607.11625} {\path{arXiv:2607.11625}}.

\bibitem{HotQCD:2012vvd}
A.~Bazavov, et~al., {The chiral transition and $U(1)_A$ symmetry restoration
  from lattice QCD using Domain Wall Fermions}, Phys. Rev. D 86 (2012) 094503.
\newblock \href {http://arxiv.org/abs/1205.3535} {\path{arXiv:1205.3535}},
  \href {https://doi.org/10.1103/PhysRevD.86.094503}
  {\path{doi:10.1103/PhysRevD.86.094503}}.

\bibitem{Buchoff:2013nra}
M.~I. Buchoff, M.~Cheng, N.~H. Christ, H.~T. Ding, C.~Jung, et~al., {The QCD
  chiral transition, $U(1)_A$ symmetry and the Dirac spectrum using domain wall
  fermions}, Phys. Rev. D 89~(5) (2014) 054514.
\newblock \href {http://arxiv.org/abs/1309.4149} {\path{arXiv:1309.4149}},
  \href {https://doi.org/10.1103/PhysRevD.89.054514}
  {\path{doi:10.1103/PhysRevD.89.054514}}.

\bibitem{Ding:2020hxw}
H.~T. Ding, S.~T. Li, A.~Tomiya, X.~D. Wang, Y.~Zhang, {Chiral properties of
  (2+1)-flavor QCD in strong magnetic fields at zero temperature}, Phys. Rev. D
  104~(1) (2021) 014505.
\newblock \href {http://arxiv.org/abs/2008.00493} {\path{arXiv:2008.00493}},
  \href {https://doi.org/10.1103/PhysRevD.104.014505}
  {\path{doi:10.1103/PhysRevD.104.014505}}.

\bibitem{Kilcup:1986dg}
G.~W. Kilcup, S.~R. Sharpe, {A Tool Kit for Staggered Fermions}, Nucl. Phys. B
  283 (1987) 493--550.
\newblock \href {https://doi.org/10.1016/0550-3213(87)90285-9}
  {\path{doi:10.1016/0550-3213(87)90285-9}}.

\bibitem{Ding:2021cwv}
H.~T. Ding, S.~T. Li, Q.~Shi, X.~D. Wang, {Fluctuations and correlations of net
  baryon number, electric charge and strangeness in a background magnetic
  field}, Eur. Phys. J. A 57~(6) (2021) 202.
\newblock \href {http://arxiv.org/abs/2104.06843} {\path{arXiv:2104.06843}},
  \href {https://doi.org/10.1140/epja/s10050-021-00519-3}
  {\path{doi:10.1140/epja/s10050-021-00519-3}}.

\bibitem{Follana:2006rc}
E.~Follana, Q.~Mason, C.~Davies, K.~Hornbostel, G.~P. Lepage, J.~Shigemitsu,
  H.~Trottier, K.~Wong, {Highly improved staggered quarks on the lattice, with
  applications to charm physics}, Phys. Rev. D 75 (2007) 054502.
\newblock \href {http://arxiv.org/abs/hep-lat/0610092}
  {\path{arXiv:hep-lat/0610092}}, \href
  {https://doi.org/10.1103/PhysRevD.75.054502}
  {\path{doi:10.1103/PhysRevD.75.054502}}.

\bibitem{DElia:2010abb}
M.~D'Elia, S.~Mukherjee, F.~Sanfilippo, {QCD Phase Transition in a Strong
  Magnetic Background}, Phys. Rev. D 82 (2010) 051501.
\newblock \href {http://arxiv.org/abs/1005.5365} {\path{arXiv:1005.5365}},
  \href {https://doi.org/10.1103/PhysRevD.82.051501}
  {\path{doi:10.1103/PhysRevD.82.051501}}.

\bibitem{Smit:1986fn}
J.~Smit, J.~C. Vink, {Remnants of the Index Theorem on the Lattice}, Nucl.
  Phys. B 286 (1987) 485--508.
\newblock \href {https://doi.org/10.1016/0550-3213(87)90451-2}
  {\path{doi:10.1016/0550-3213(87)90451-2}}.

\bibitem{Al-Hashimi:2008quu}
M.~H. Al-Hashimi, U.~J. Wiese, {Discrete Accidental Symmetry for a Particle in
  a Constant Magnetic Field on a Torus}, Annals Phys. 324 (2009) 343--360.
\newblock \href {http://arxiv.org/abs/0807.0630} {\path{arXiv:0807.0630}},
  \href {https://doi.org/10.1016/j.aop.2008.07.006}
  {\path{doi:10.1016/j.aop.2008.07.006}}.

\end{thebibliography}
\end{document}